\documentclass[final,3p,twocolumn]{elsarticle}

\usepackage[utf8]{inputenc}
\usepackage[english]{babel}

\makeatletter\adddialect\l@English\l@english\makeatother
\usepackage{tikz}
\usepackage{paralist}
\usepackage{subcaption} \usepackage{graphicx}
\usepackage{amsmath} \usepackage{amssymb}
\usepackage{xcolor}
\usepackage{natbib}

\graphicspath{{./fig/}}

\begin{document} 
\begin{frontmatter}
\title{The Neutrino Opacity of Neutron Rich Matter}

\author{P. N. Alcain and  C. O. Dorso}

\address{Departamento de F\'isica, FCEyN, UBA and IFIBA,
  Conicet, Pabell\'on 1, Ciudad Universitaria, 1428 Buenos
  Aires, Argentina} \address{IFIBA-CONICET}

\date{\today} 

\begin{abstract}
  The study of neutron rich matter, present in neutron star,
  proto-neutron stars and core-collapse supernovae, can lead to
  further understanding of the behavior of nuclear matter in highly
  asymmetric nuclei. Heterogeneous structures are expected to exist in
  these systems, often referred to as nuclear pasta. We have carried
  out a systematic study of neutrino opacity for different
  thermodynamic conditions in order to assess the impact that the
  structure has on it. We studied the dynamics of the neutrino opacity
  of the heterogeneous matter at different thermodynamic conditions
  with semiclassical molecular dynamics model already used to study
  nuclear multifragmentation. For different densities, proton
  fractions and temperature, we calculate the very long range opacity
  and the cluster distribution. The neutrino opacity is of crucial
  importance for the evolution of the core-collapse supernovae and the
  neutrino scattering.
\end{abstract}
\end{frontmatter}

\section{Introduction}\label{sc:intro}
Most neutron stars are supernovae remnants, that happen when the hot
and dense iron core of a dying massive star collapses. This gives rise
to a system known as \emph{proto-neutron star}, which eventually ends
up in a neutron star. During the collapse, several nuclear processes
take place in the inner core of the star --- electron capture,
photodisintegration, Urca, etc.  Neutrinos are copiously produced in
core collapse supernovae, proto-neutron star and, to a lesser extent,
in the core of neutron stars. These neutrinos flow outwards, and their
emission is the main mean by which the proto-neutron stars cool
down. Therefore, the interaction between the neutrinos and
heterogeneous neutron rich matter is key to comprehend the dynamics of
the systems under study.

Several models have been developed to study nuclear pasta, and they
have shown that these structures arise due to the interplay between
nuclear and Coulomb forces in an infinite medium. Nevertheless, the
dependence of the observables in different thermodynamic conditions
has not been studied in depth. The original works of Ravenhall
\emph{et al.}~\cite{ravenhall_structure_1983} and Hashimoto \emph{et
  al.}~\cite{hashimoto_shape_1984} used a compressible liquid drop
model, and have shown that the now known as the \emph{pasta phases}
--\emph{lasagna}, \emph{spaghetti} and \emph{gnocchi}-- are solutions
to the ground state of neutron star matter. From then on, different
approaches have been taken, that we roughly classify in two
categories: mean field or microscopic.

Mean field works include the Liquid Drop Model, by Lattimer \emph{et
  al.}~\cite{page_minimal_2004}, Thomas-Fermi, by Williams and
Koonin~\cite{williams_sub-saturation_1985}, among
others~\cite{oyamatsu_nuclear_1993, lorenz_neutron_1993,
  cheng_properties_1997, watanabe_thermodynamic_2000,
  watanabe_electron_2003, nakazato_gyroid_2009}. Microscopic models
include Quantum Molecular Dynamics, used by Maruyama \emph{et
  al.}~\cite{maruyama_quantum_1998, kido_md_2000} and by
Watanabe~\emph{et al.}\cite{watanabe_structure_2003}, Simple
Semiclassical Potential, by Horowitz~\emph{et
  al.}~\cite{horowitz_nonuniform_2004} and Classical Molecular
Dynamics, used in our previous works~\cite{dorso_topological_2012}.

In some recent studies, phases different from the typical
\emph{nuclear pasta} were found. The work by Nakazato \emph{et
  al.}~\cite{nakazato_gyroid_2009}, inspired by polymer systems, found
also gyroid and double-diamond structures, with a compressible liquid
drop model. Dorso \emph{et al.}~\cite{dorso_topological_2012} obtained
pasta phases different from those already mentioned with molecular
dynamics, studying mostly their characterization at very low
temperatures. In our previous work~\cite{alcain_beyond_2014} we have
shown that these new pasta phases had an opacity peak (i.\ e., a local
maximum in the opacity) in the characteristic wavelength of the Urca
neutrinos for symmetrical neutron star matter. We will refer to all
these different non homogeneous phases as \emph{Generalized Nuclear
  Pasta} (GNP).

Among the advantages of classical or semiclassical models are the
accessibility to position and momentum of all particles at all times,
which allows the calculation of correlations of all orders. Moreover,
no specific structure is hardcoded in the model, as it happens with
most mean field models. This enables the study of the structure of the
nuclear medium from a particle-wise point of view. Many models exist
with this goal, including quantum molecular
dynamics~\cite{maruyama_quantum_1998}, simple-semiclassical
potential~\cite{horowitz_nonuniform_2004} and classical molecular
dynamics~\cite{lenk_accuracy_1990}. In these models the Pauli
repulsion between nucleons of equal isospin is hard-coded in the
interaction. On the other hand, a specific Pauli potential developed
in~\cite{dorso_classical_1987} was used in the
QCNM~\cite{dorso_classical_1988} and later in
Ref.~\cite{hartnack_quantum_1989}.

In the works done by Horowitz \emph{et
  al.}~\cite{horowitz_neutrino-pasta_2004, horowitz_nonuniform_2004},
the neutrino opacity and mean free path was calculated for a
specific temperature and proton fraction. With these results they
showed that a very long range structure (\emph{nuclear pasta}) emerges
in calculations using models with long-range Debye-like repulsion and
short-range nuclear-like interaction. For the studied system, this
very long range structure has an opacity peak in the energy region
of Urca neutrinos for the very diluted \emph{gnocchi} phase.

We build the present work upon this result, for a different
microscopic model with the same qualitative characteristics, also
extending the studied thermodynamic region for different proton
fractions, temperatures and densities. We calculate \emph{i)} the
opacity for long wavelengths compared to the interparticle distance of
nuclear matter ($r_{nn}\approx 1.8\,\text{fm}$) and \emph{ii)} the
cluster mass distribution. This later quantity allows us to determine
whether the pasta phase is finite or infinite, the characteristics of
each phase and insight into the neutron rich gas in equilibrium with
it.

In Section~\ref{sc:model} we introduce the model used along this work,
that includes the potential parametrization (\ref{ssc:cmd}), the
Coulomb interaction (\ref{ssc:coulomb}) and magnitudes of interest
(\ref{ssc:moi}): cluster distribution and neutrino
opacity. Section~\ref{ssc:clusters} shows the cluster distribution for
different configurations, and in Section~\ref{ssc:opacity} we study in
greater detail the opacity of the pasta to long wavelength neutrinos,
for different thermodynamic parameters. Finally, we draw conclusions
in Section~\ref{sc:conc}. In the appendix, a detailed analysis of the
static structure factor calculation is performed.

\section{The Model}\label{sc:model}
\subsection{Classical Molecular Dynamics}\label{ssc:cmd}
In this work, we study GNP with the classical molecular dynamics model
CMD.\@It has been used in several heavy-ion reaction studies to: help
understand experimental data~\cite{chernomoretz_quasiclassical_2002};
identify phase-transition signals and other critical
phenomena~\cite{lopez_lectures_2000, barranon_searching_2001,
  dorso_selection_2001, barranon_critical_2003, barranon_time_2007};
and explore the caloric curve~\cite{barranon_entropy_2004} and
isoscaling~\cite{dorso_dynamical_2006, dorso_isoscaling_2011}. CMD
uses two two-body potentials to describe the interaction of nucleons,
which are a combination of Yukawa potentials:
\begin{align}
  V^{\text{CMD}}_{np}(r) &=v_{r}\exp(-\mu_{r}r)/{r}-v_{a}\exp(-\mu_{a}r)/{r}\\ 
  V^{\text{CMD}}_{nn}(r) &=v_{0}\exp(-\mu_{0}r)/{r}
\end{align}
where $V_{np}$ is the potential between a neutron and a proton, and
$V_{nn}$ is the repulsive interaction between either $nn$ or $pp$. The
cutoff radius is $r_c=5.4\,\text{fm}$ and for $r>r_c$ both potentials
are set to zero. The Yukawa parameters $\mu_r$, $\mu_a$ and $\mu_0$
were determined to yield an equilibrium density of $\rho_0=0.16
\,\text{fm}^{-3}$, a binding energy $E(\rho_0)=16
\,\text{MeV/nucleon}$ and a compressibility of $250\,\text{MeV}$.

To simulate an infinite medium, we used this potential with $N = 5500$
particles under periodic boundary conditions, with different proton
fraction (i.\ e.\ with $x = Z/A = 0.1 < x < 0.5$) in cubical boxes with sizes
adjusted to have densities between $\rho=0.001 \,\text{fm}^{-3} \le
\rho \le 0.08\,\text{fm}^{-3}$. This simulations have been done with
LAMMPS~\cite{plimpton_fast_1995}, using its GPU
package~\cite{brown_implementing_2012}.

\subsection{Coulomb interaction in the model}\label{ssc:coulomb}

Since a neutralizing electron gas embeds the nucleons in the neutron
star crust, the Coulomb forces among protons are screened. We model
this screening effect with the Thomas-Fermi approximation, used with
various nuclear models~\cite{maruyama_quantum_1998,
  dorso_topological_2012, horowitz_neutrino-pasta_2004}. According to
this approximation, protons interact via a Yukawa-like potential, with
a screening length $\lambda$:
\begin{equation}
 V_{TF}(r) = q^2\frac{e^{-r/\lambda}}{r}.
\end{equation}

Theoretical estimates for the screening length $\lambda$ are
$\lambda\sim100\,\text{fm}$~\cite{fetter_quantum_2003}, but we set the
screening length to $\lambda=20\,\text{fm}$. This choice was based on
previous studies~\cite{alcain_effect_2014}, where we have shown that
this value is enough to adequately reproduce the expected length scale
of density fluctuations for this model, while larger screening lengths
would be a computational difficulty. We analyze the opacity to
neutrinos of the structures for different proton fractions and
densities.

\subsection{Magnitudes of Interest}\label{ssc:moi}

\subsubsection{Neutrino Opacity}\label{ssc:moi_no}
Neutron rich matter is a neutral system composed of a neutron enriched
mixture of neutrons and protons embedded in a degenerate electron
gas. This kind of matter can develop heterogeneous structures usually
referred to as \emph{nuclear pasta}. As seen in
Ref.~\cite{horowitz_nonuniform_2004, horowitz_neutrino-pasta_2004},
the neutron-neutron static structure factor $S(q)$ of the nuclear
pasta describes coherent neutrino scattering. This phenomenon is
expected to dominate the neutrino opacity for certain wavelengths. The
scattering cross section is related to the static structure factor
through

\begin{equation}
 \sigma_{\text{total}} = \sigma_{\text{free neutron}} \times S(q)
\end{equation}
The neutrino scattering cross section of a free neutron is given by:
\begin{equation}
\sigma_{\text{free neutron}} = \frac{G_F^2E_\nu^2}{6\pi}
\end{equation}
with $G_F$ the Fermi coupling and $E_\nu$ the energy of the
neutrino. With this in mind, the cross section is:
\begin{equation}
\sigma_{\text{total}} = \frac{G_F^2E_\nu^2}{6\pi}\, S(q)
\label{eq:opac}
\end{equation}
Since the neutrino mass ($m_\nu \approx 10^{-2}\,\text{eV}$) is
negligible for energies in the MeV range, the relation between
the energy and the wave number is $E_\nu = \hbar q$.

To find the opacity of heterogeneous matter we calculated the
structure factor of the system for a broad range of wavelengths of
interest related to the pasta structure, and then searched for the
maximum. To have an idea of the mass distribution of the system we
calculate the pair distribution function of the neutrons $g_{nn}(r)$,
which is related to the average number of neutrons at a distance $r$
away from a given neutron, in a shell of thickness $\text{d}r$:

\begin{equation}
\text{d}n = \frac{N}{V} g_{nn}(r) 4\pi r^2 \text{d}r
\end{equation}
The structure factor $S(q)$ is the Fourier transform of the pair
distribution function: 
\begin{equation}
S_{nn}(q) = 1 + \rho \int_V{\text{d}r\, \text{e}^{-i\,q\,r} \left[g_{nn}(r)
- 1\right]}  
\end{equation}
This expression is for an angle averaged $S(q)$, since collapsing
cores are polycrystalline, and the orientation of each grain of the
crystal is random~\cite{sonoda_impact_2007}.

Since there is a transition from infinite clusters (totally connected
structures) to finite clusters (usually small spherical clusters of
neutron star matter), dependent on the density and the temperature, it
is of relevance to relate the neutrino opacity with the cluster
structure of the system. The finite pasta, \emph{gnocchi}, exists for
densities below a given threshold and, as shown below, it is the
finite pasta that accounts for the opacity for long wavelengths.

\subsubsection{Cluster recognition: Identifying pasta phases}

In typical configurations we have not only the structure known as
nuclear pasta, but also a nucleon gas that surrounds the nuclear
pasta. In order to properly characterize the pasta phases, we must
identify which atoms belong to the pasta phases and which belong to
this gas. To do so, we have to find the clusters that are formed along
the simulation.

One of the algorithms to identify cluster formation is Minimum
Spanning Tree (MST). In MST algorithm, two particles belong to the
same cluster $\{C^{\text{MST}}_n\}$ if the relative distance between
the particles is less than a cutoff distance $r_{cut}$:
\begin{equation}
  i \in C^{\text{MST}}_n \Leftrightarrow \exists j \in C_n \mid
  r_{ij} < r_{cut}
\end{equation}

This cluster definition works correctly for systems in which relative
velocities between particles are not relevant (for example, the
asymptotic state of a fragmenting nucleus), and it is based on the
attractive tail of the nuclear interaction. However, if the system has
a high temperature, we can have two particles that are closer than the
cutoff radius, but with a large relative kinetic energy.

To deal with situations of non-zero temperatures, we need to take into
account the relative momentum among particles. One of the most
sophisticated methods to accomplish this is the Early Cluster
Recognition Algorithm (ECRA)~\cite{dorso_early_1993}. In this
algorithm, the particles are partitioned in different disjoint
clusters $C^{\text{ECRA}}_n$, with the total energy in each cluster:
\begin{equation}
  \epsilon_n = \sum_{i \in C_n} K^{CM}_i +  \sum_{i,j \in C_n} V_{ij}
\end{equation}
where $K^{CM}_i$ is the kinetic energy relative to the center of mass
of the cluster and $V_{ij}$ is the interaction potential energy
between particles $i$ and $j$. The set of clusters $\{C_n\}$ then is
the one that minimizes the sum of all the cluster energies
$E_{\text{partition}} = \sum_n \epsilon_n$.

\begin{figure}
  \centering
  \includegraphics[width=\columnwidth]{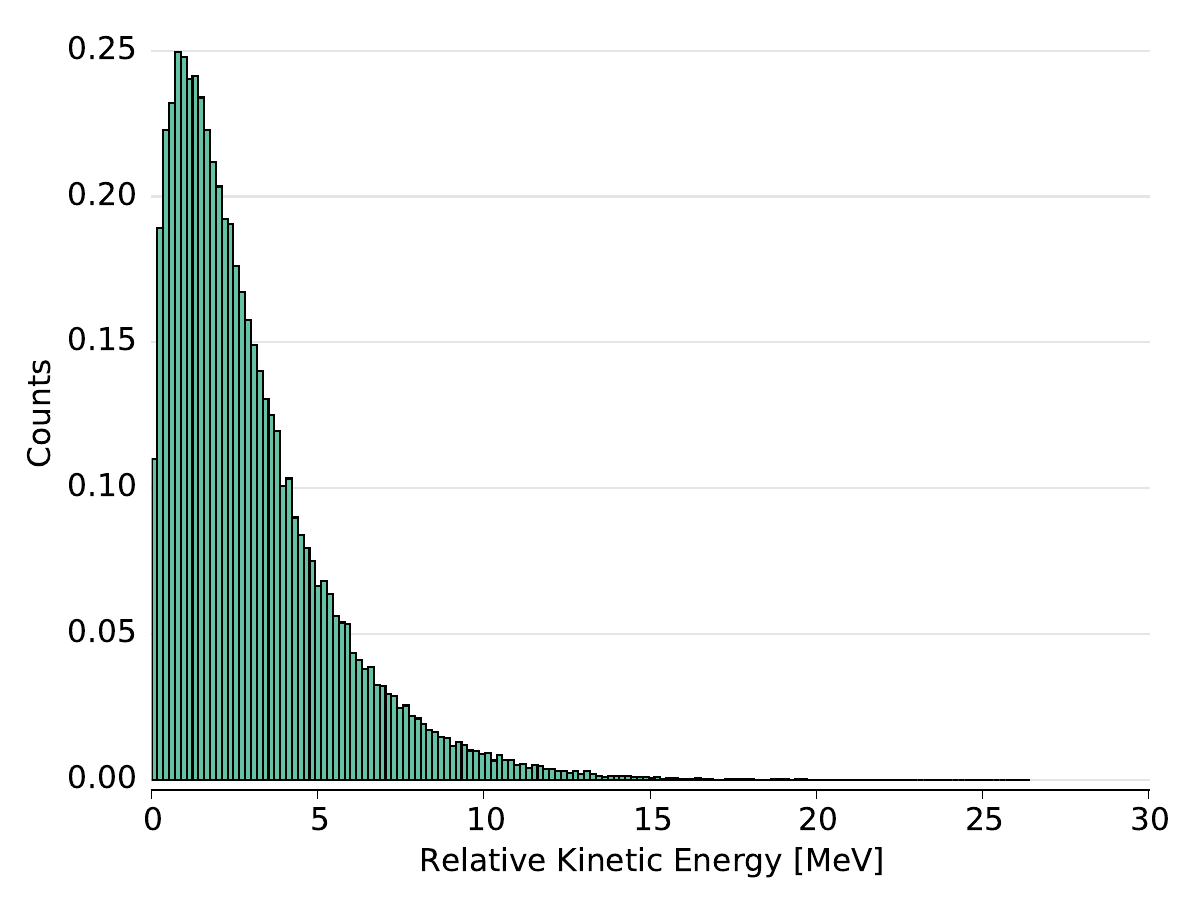}
  \caption{(Color online) Relative kinetic energy for pairs inside MST
    clusters.}
\label{fig:kinetic}
\end{figure}

ECRA algorithm can be easily used for small
systems~\cite{dorso_fluctuation_1994}, but being a combinatorial
optimization, it cannot be used in large systems. While finding ECRA
clusters is very expensive computationally, using simply MST clusters
can lead to results extremely biased in favor of large clusters. We have
decided to go for a middle ground choice, the Minimum Spanning Tree
Energy (MSTE) algorithm~\cite{dorso_topological_2012}. This algorithm
is a modification of MST, taking into account the kinetic
energy. According to MSTE, two particles belong to the same cluster
$\{C^{\text{MSTE}}_n\}$ if they are energy bound:
\begin{equation}
  i \in C^{\text{MSTE}}_n \Leftrightarrow \exists j \in C_n :
  V_{ij}+ K_{ij} \le 0
\end{equation}
While this algorithm doesn't yield the same theoretically sound
results from ECRA, it still avoids the largest pitfall of naïve MST
implementations for the temperatures used in this work. To illustrate
this concept, we show in figure~\ref{fig:kinetic} the relative
kinetic energy of pairs that are bound by MST algorithm, with $r_{cut}
= 5.4\,\text{fm}$, for a system with $x = 0.5$, $\rho =
0.04\,\text{fm}^{-3}$, $T = 1.0\,\text{MeV}$. We can see that a
considerable amount of pairs have a relative energy larger than
$5\,\text{MeV}$. 

\begin{figure}
  \centering
  \includegraphics[width=\columnwidth]{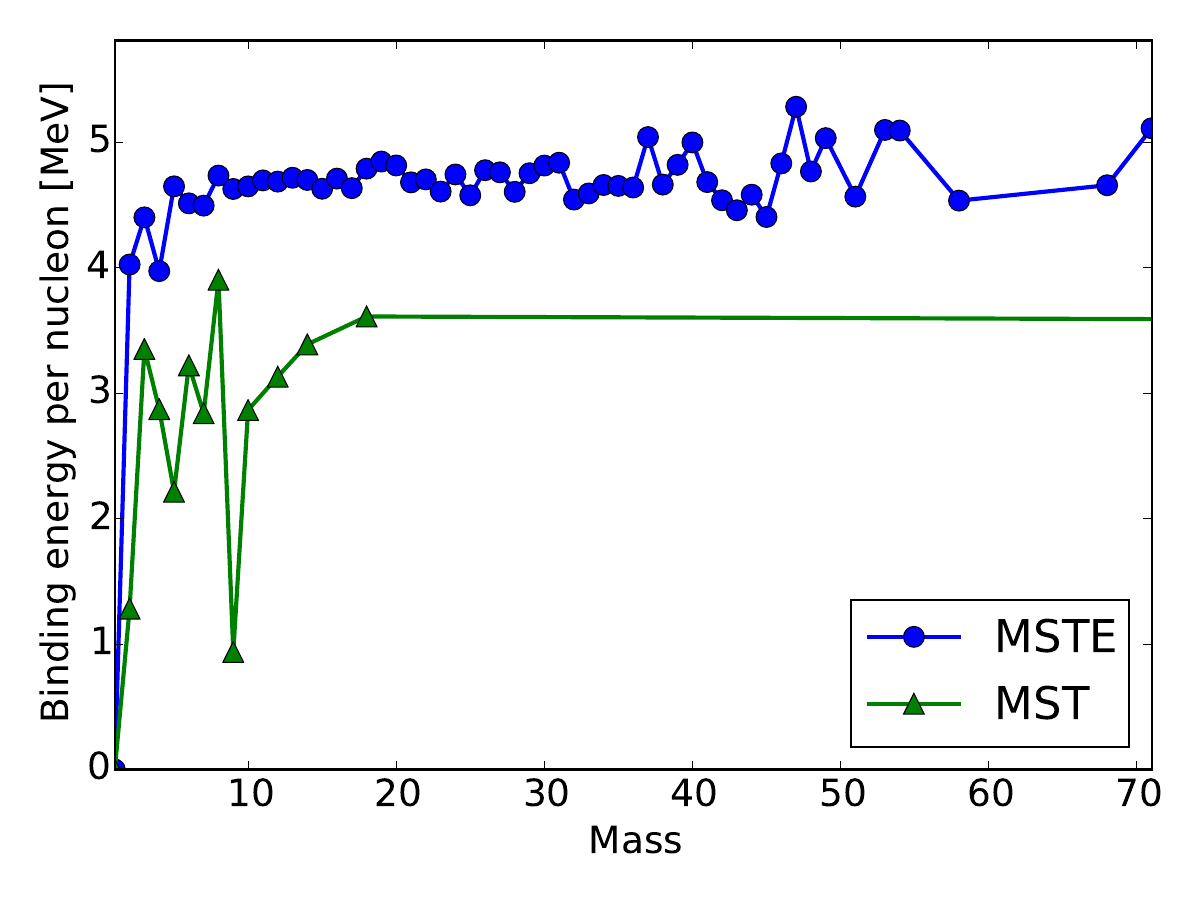}
  \caption{(Color online) Binding energy for MST and MSTE clusters. We
    can see that for every cluster size, MSTE clusters are more bound
    than MST ones.}
\label{fig:energy-clusters}
\end{figure}

Even further, for systems of density $\rho = 0.01\,\text{fm}^{-3}$ and
proton fraction $x = 0.3$ with the lowest temperature studied ($T =
0.5\,\text{MeV}$), we tallied the binding energy per nucleon $E_B$ for
the different cluster mass (this is related to the $\epsilon_n$ from
the ECRA definition: $E_B = -\epsilon_n/m_n$, with $m$ the mass of the
cluster $C_n$) that appeared along the different snapshots. This was
done both for MST and MSTE clusters, and the obtained results are in
figure~\ref{fig:energy-clusters}. In this figure we can see that for
every cluster size, MSTE clusters have a larger binding energy than
MST clusters.

\section{Results}\label{results}
\subsection{Clusters}\label{ssc:clusters}

\begin{figure*}  \centering
  \begin{subfigure}[h!]{0.80\columnwidth}
    \includegraphics[width=\columnwidth]{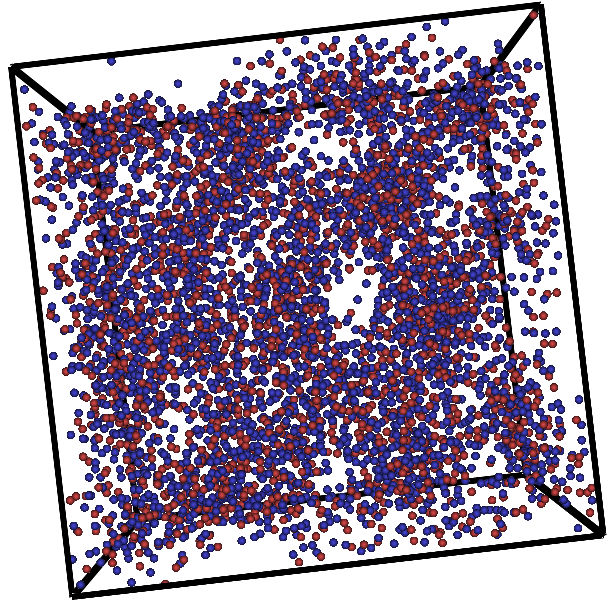}
    \caption{$x=0.4,\, T=0.5\,\text{MeV}$}
\label{subfig:04-05}
  \end{subfigure}
  \begin{subfigure}[h!]{0.80\columnwidth}
    \includegraphics[width=\columnwidth]{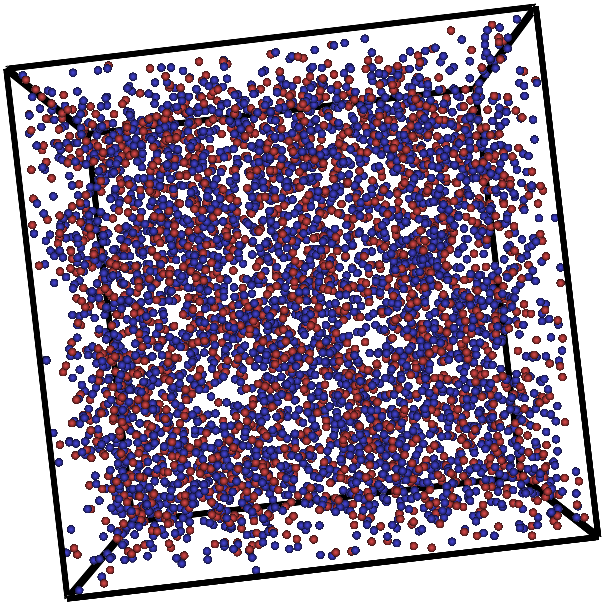}
    \caption{$x=0.4,\, T=1.0\,\text{MeV}$}
\label{subfig:04-10}
  \end{subfigure}
  \begin{subfigure}[h!]{0.80\columnwidth}
    \includegraphics[width=\columnwidth]{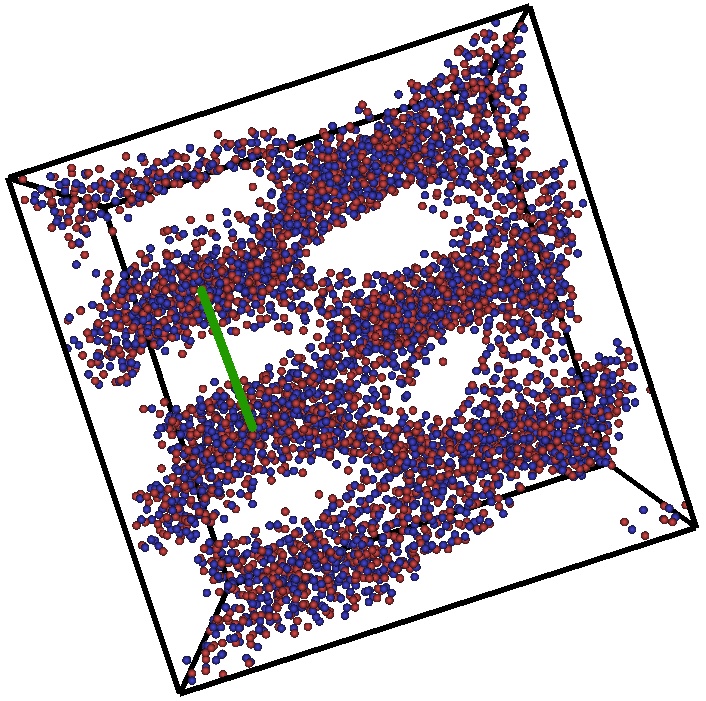}
    \caption{$x=0.5,\, T=0.5\,\text{MeV}$}
\label{subfig:05-05}
  \end{subfigure}
  \begin{subfigure}[h!]{0.80\columnwidth}
    \includegraphics[width=\columnwidth]{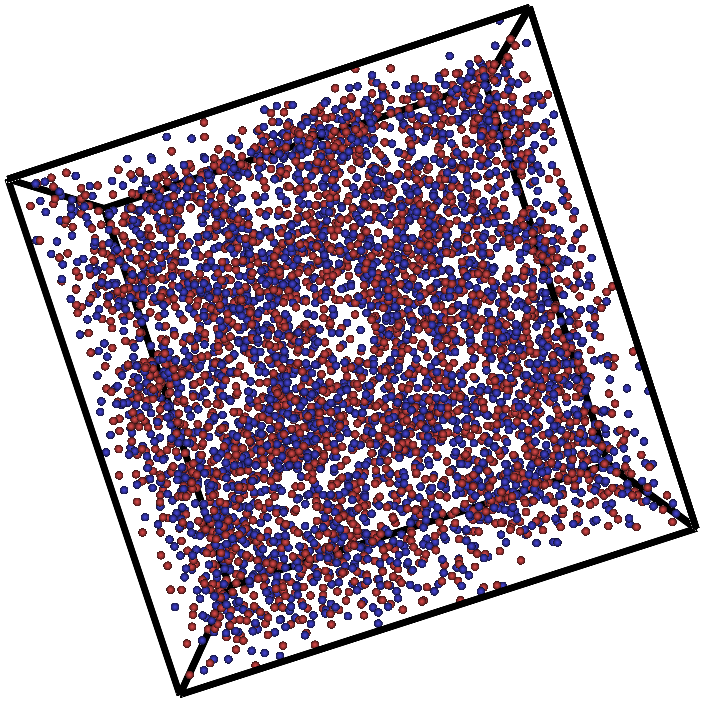}
    \caption{$x=0.5,\, T=1.0\,\text{MeV}$}
\label{subfig:05-10}
  \end{subfigure}
  \caption{(Color online) Snapshots of a system with density $\rho =
    0.04\,\text{fm}^{-3}$ for different values of proton fraction and
    temperature, generated with
    VisIt~\cite{childs_contract-based_2005}. Structures obtained at
    $T=0.5\,\text{MeV}$ differ substantially. Nevertheless both show
    inhomogeneities. We can see in panel~\ref{subfig:05-05} a green
    line marking a correlation length of $\approx 15\,\text{fm}$.}
\label{fig:morpho}
\end{figure*}

\begin{figure*}
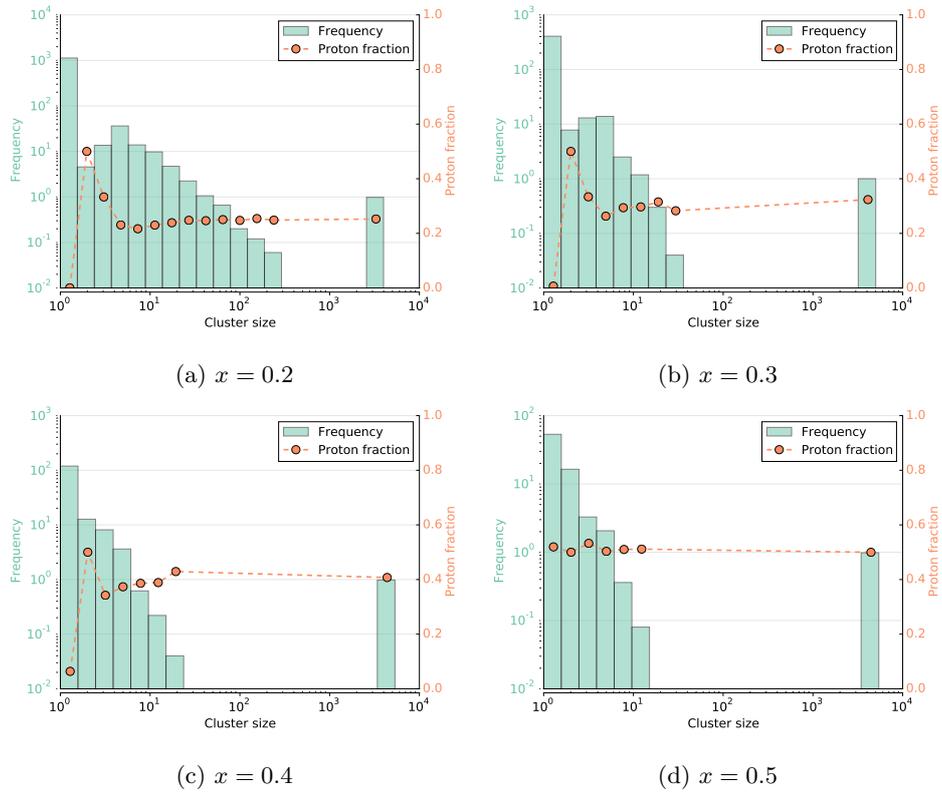
  \centering
  \begin{subfigure}[h!]{0.8\columnwidth}
    \includegraphics[width=\columnwidth]{{{mste_0.2_0.04_2.0}}}
    \caption{$x=0.2$}
  \end{subfigure}
  \begin{subfigure}[h!]{0.8\columnwidth}
    \includegraphics[width=\columnwidth]{{{mste_0.3_0.04_2.0}}}
    \caption{$x=0.3$}
  \end{subfigure}
  \begin{subfigure}[h!]{0.8\columnwidth}
    \includegraphics[width=\columnwidth]{{{mste_0.4_0.04_2.0}}}
    \caption{$x=0.4$}
  \end{subfigure}
  \begin{subfigure}[h!]{0.8\columnwidth}
    \includegraphics[width=\columnwidth]{{{mste_0.5_0.04_2.0}}}
    \caption{$x=0.5$}
  \end{subfigure} 
  \caption{(Color online) Cluster distribution with MSTE algorithm for
    temperature $T = 2.0\,\text{MeV}$, density $\rho =
    0.04\,\text{fm}^{-3}$ and different proton fractions. For the
    lowest of the studied proton fractions, $x = 0.2$, the large
    cluster has a higher proton fraction (about $30\%$ higher) and
    there are many isolated neutrons. Please note that the scales are
    different for each graph.}
\label{fig:cluster}
\end{figure*}

In figure~\ref{fig:morpho} we show four different snapshots for proton
fractions of $x=0.4$ and $x=0.5$ and temperature $T=0.5\,\text{MeV}$
and $T=1.0\,\text{MeV}$. We clearly see that the structures are no
longer limited to those originally proposed by Ravenhall \emph{et
  al.}~\cite{ravenhall_structure_1983}. To study them further we can
see in figure~\ref{fig:cluster} the corresponding cluster distribution
according to MSTE algorithm. In this figure, we can see that for a
proton fraction $x=0.2$ there are many isolated nucleons that are
almost exclusively neutrons. These form the previously mentioned
neutron gas that embeds the underlying proton structure.

\begin{figure}
  \centering
  \includegraphics[width=\columnwidth]{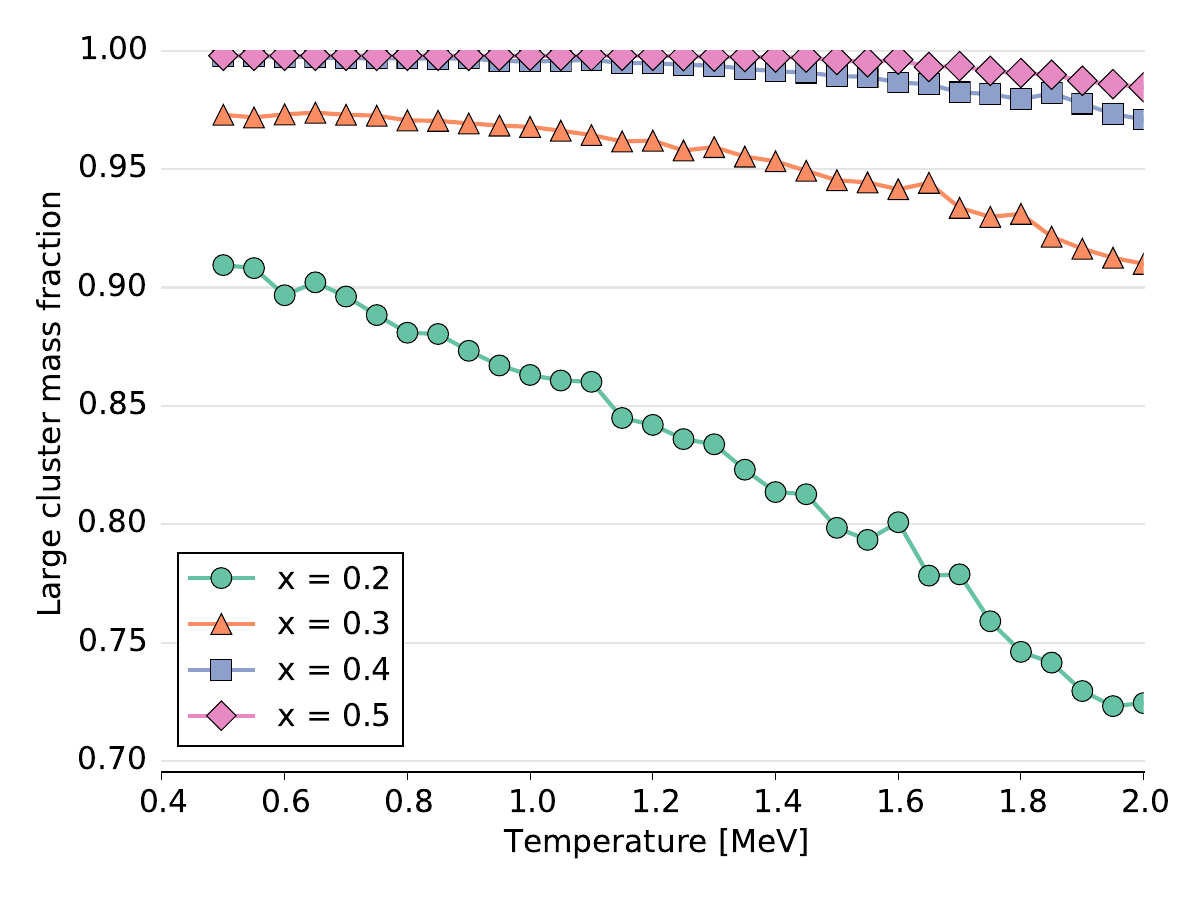}
  \caption{(Color online) Mass fraction of the largest cluster for
    $\rho = 0.04\,\text{fm}^{-3}$ for different values of $x$.}
\label{fig:large_mass}
\end{figure}

Another consequence of the neutron gas is that the proton fraction of
the GNP structure is slightly higher than the proton fraction in the
simulation cell. We can see from figure~\ref{fig:cluster} that the
proton fraction in the large cluster is about $x = 0.24$, while the
macroscopic proton fraction is $x = 0.2$. In
figure~\ref{fig:large_mass} we show the mass fraction of the largest
cluster in terms of the temperature and the proton fraction, and note
that even for very high temperatures ($T = 2.0\,\text{MeV}$) a large
cluster appears for every proton fraction. In particular, the smallest
of the largest clusters contains more than $50\%$ of the total mass of
the system.

\subsection{Neutrino Opacity}\label{ssc:opacity}

As explained in Section~\ref{ssc:moi_no}, we calculate the neutrino
opacity of the neutron rich matter.  Figure~\ref{fig:gr_sq_gnocchi}
shows the pair distribution function, structure factor (see appendix
for a detailed explanation of its calculation) and opacity for the
\emph{gnocchi} phase. In the pair distribution function we can
identify (marked with $\color{green} \blacktriangledown$) the peak
that corresponds to the crystalline structure of the nucleons within
the pasta --- neutron correlation with nearest neighbors ---, and also
a very long range order (marked with a dashed line
$\color{red}-\,-$.); this interaction leads to the peak for low
wavenumbers in the structure factor, related to the pasta
structures. The structure factor displays a pasta peak (see
Figure~\ref{fig:gr_sq_gnocchi}) located at
$q_\text{peak} = 0.37\,\text{fm}^{-1}$ (that translates to a neutrino
energy of $E_\nu \approx 70\,\text{MeV}$) for this \emph{gnocchi}
phase and with a full width at half maximum of about
$\Delta q_{\text{FWHM}} = 0.08\,\text{fm}^{-1}$
($\Delta E_{\text{FWHM}} \approx 15\,\text{MeV}$), by so defining a
range of wavelengths in which the structure is considerably opaque.


\begin{figure}  \centering
  \begin{subfigure}[h!]{\columnwidth}
    \includegraphics[width=\columnwidth]{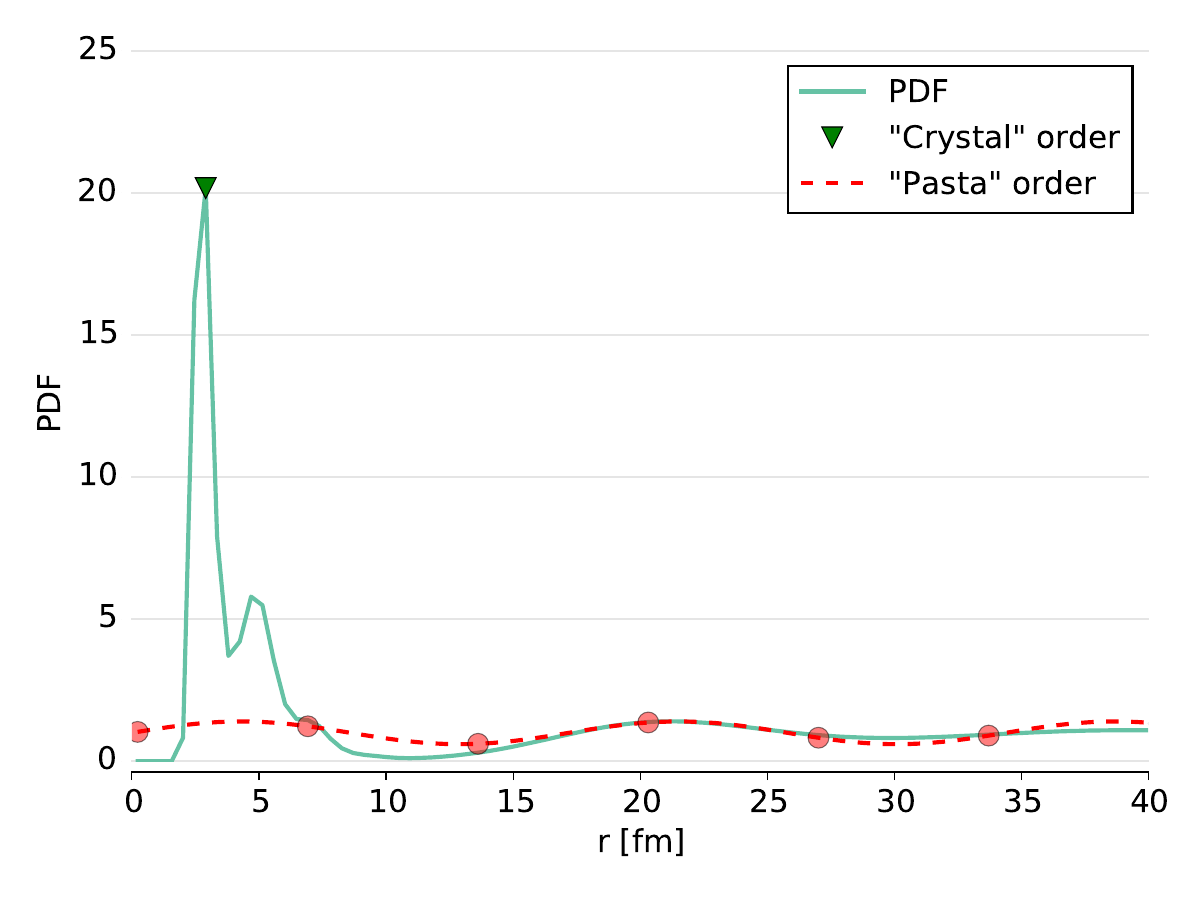}
    \caption{Pair distribution function.}
\label{sfig:gr_gnocchi}
  \end{subfigure}
  \begin{subfigure}[h!]{\columnwidth}
    \includegraphics[width=\columnwidth]{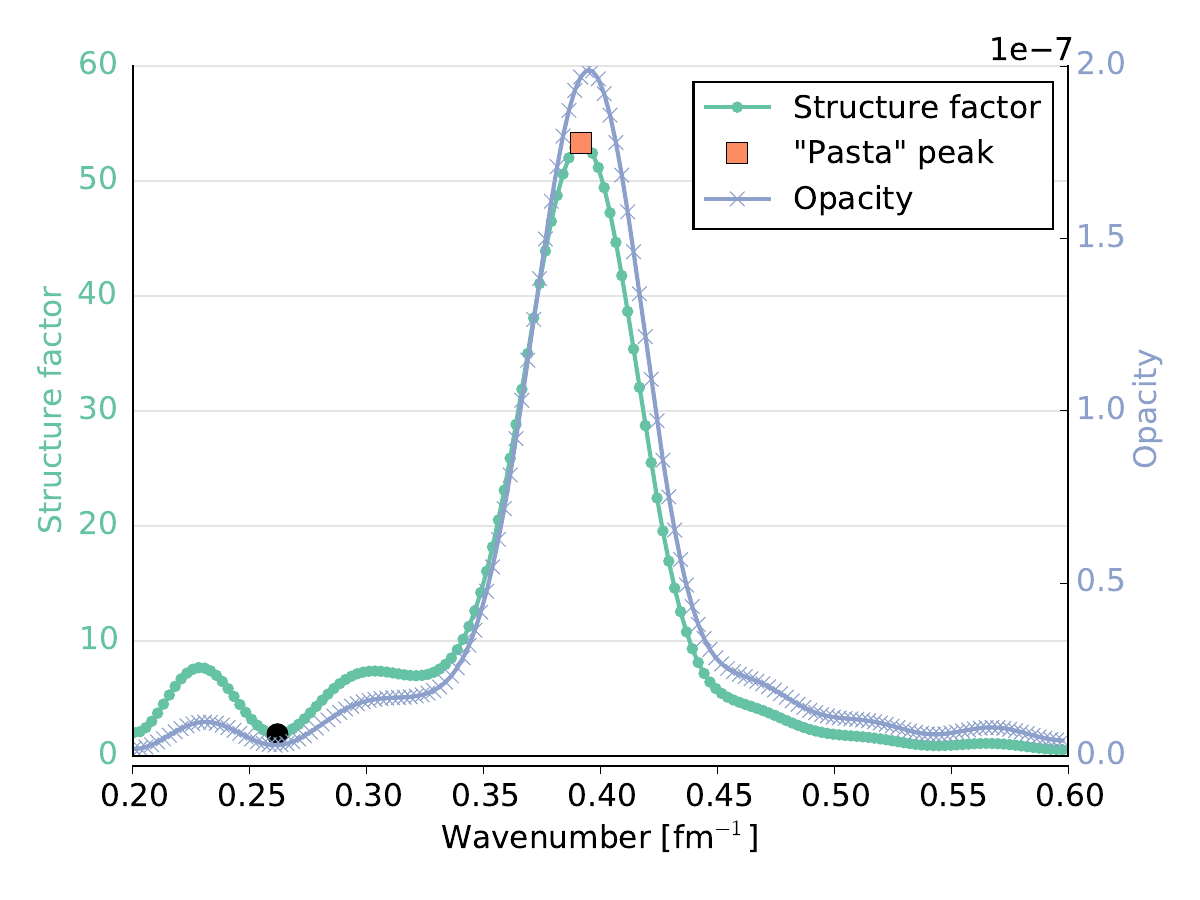}
    \caption{Static structure factor.}
\label{sfig:ssf_gnocchi}
  \end{subfigure}
  \caption{(Color online)~\ref{sfig:gr_gnocchi} Pair distribution
    function; and~\ref{sfig:ssf_gnocchi} static structure factor and
    opacity for a system with proton fraction $x=0.4$, density
    $\rho=0.01\,\text{fm}^{-3}$ and temperature
    $T=0.5\,\text{MeV}$. The first peak in the $g(r)$ due
    to crystalline structures is marked with $\color{green}
    \blacktriangledown$, while the very long range order is marked
    with a dashed line $\color{red}-\,-$. In the structure factor we
    can see the peak located at $q_\text{peak} = 0.37\,\text{fm}^{-1}$
    with a width of about $\text{FWHM} = 0.08\,\text{fm}^{-1}$. The
    ripples for low wavenumbers are due to finite size effects.}
\label{fig:gr_sq_gnocchi}
\end{figure}

We simulated the system for a total of about 1000 different
configurations (4 different proton fractions, 10 different densities
and 30 different temperatures). For each configuration of given proton
fraction, density and temperature, we calculate the structure factor
and calculate the corresponding opacity, according to
equation~\eqref{eq:opac} and extract its maximum value for long
wavelengths. We will refer to this value as \emph{opacity peak}. A
word of caution must be said about the low temperatures. As we have
shown in a previous work~\cite{alcain_beyond_2014}, below a certain
temperature (near $1\,\text{MeV}$) the system might lock in one of
many local minima. Because of this, the system cannot be directly
simulated at low temperatures. Instead, the low temperature limit must
be obtained coming from high temperatures, carefully lowering the
temperature and checking whether the system is thermalized or not.

\begin{figure}
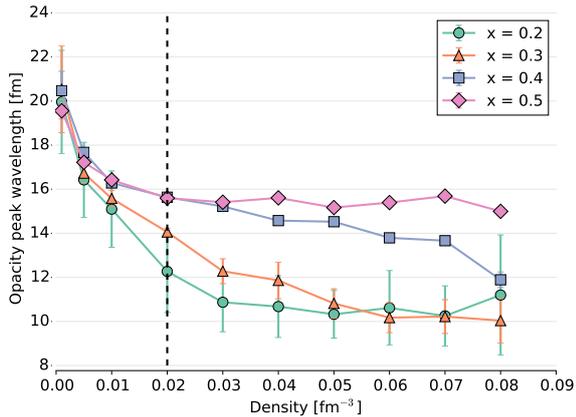
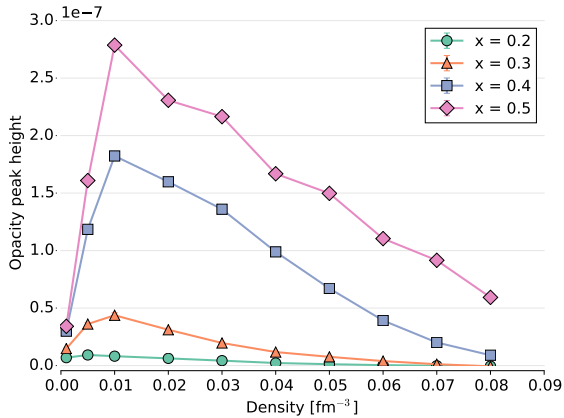
  \centering
  \begin{subfigure}[h!]{\columnwidth}
    \includegraphics[width=\columnwidth]{{{lambda_density}}}
    \caption{Opacity peak wavelength}
\label{sfig:wl}
  \end{subfigure}
  \begin{subfigure}[h!]{\columnwidth}
    \includegraphics[width=\columnwidth]{{{height_density}}}
    \caption{Opacity peak height}
\label{sfig:ht}
  \end{subfigure}
  \caption{(Color online) Opacity peak~\ref{sfig:wl} wavelength
    and~\ref{sfig:ht} height for low temperature
    ($T = 0.5\,\text{MeV}$) as a function of density for different
    proton fractions. We can see the wavelength changing rapidly for
    $\rho < 0.02\,\text{fm}^{-3}$ (\emph{gnocchi} phase) and
    stabilizing for higher densities.}
\label{fig:low_t}
\end{figure}

\begin{figure}
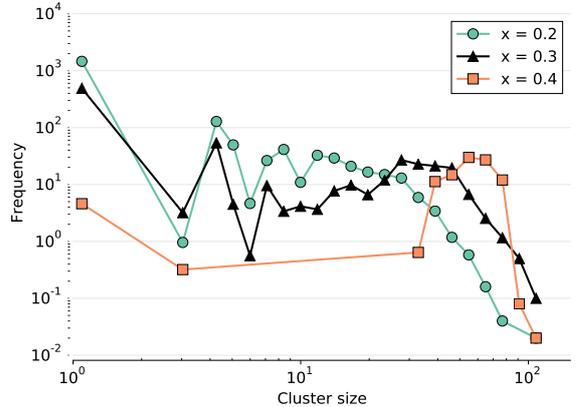
  \centering
  \includegraphics[width=\columnwidth]{{{mste_juntos}}}
  \caption{(Color online) Cluster distribution with MSTE algorithm for
    temperature $T = 0.5\,\text{MeV}$, density $\rho =
    0.01\,\text{fm}^{-3}$ and different proton fractions. We can see
    that all of them have \emph{gnocchi} mass distributions.}
\label{fig:cluster_gnocchi}
\end{figure}

Figure~\ref{fig:low_t} shows the \emph{opacity peak wavelength} and
\emph{opacity peak height} for the lowest temperature studied in this
work ($T = 0.5\,\text{MeV}$) as a function of the density. We observe
that the \emph{opacity peak wavelength} decreases as the density
increases, meaning that the correlation length of the structure
is lower as the density increases.  This is to be expected, since the
higher the density, the closer the structures are. Nevertheless, we
emphasize that the structure changes with the density, not only with
transition in morphology (e.\ g.\ from \emph{spaghetti} to
\emph{lasagna}) but also, for example, \emph{gnocchi} clusters have
different sizes for different densities. This interplay between the
structures changing internally and also changing their spatial
distribution is what results in the figure~\ref{sfig:wl}. We can see
that the \emph{opacity peak wavelength} changes rapidly for low
densities (those of \emph{gnocchi}), but tends to stabilize for the
other pasta phases. Consider also that, since the $S(q)$ has a certain
width near the peak, the structure would scatter neutrinos in a range
of wavelengths that are near said maximum. Interestingly, the
\emph{opacity peak} height reaches its maximum for
$\rho = 0.01\,\text{fm}^{-3}$, where we still have \emph{gnocchi} as
can be evidenced by the cluster distributions in
figure~\ref{fig:cluster_gnocchi}.

\begin{figure*}
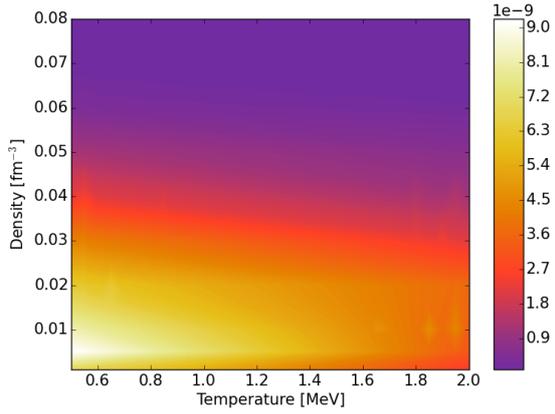
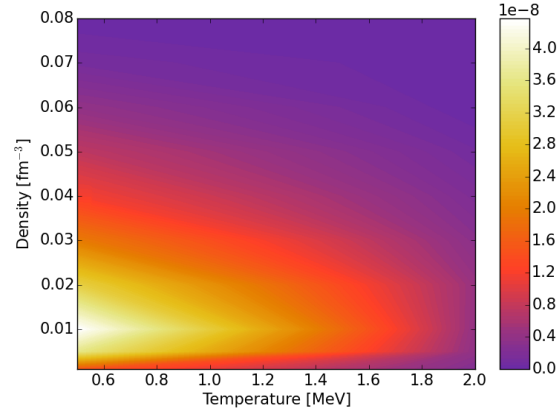
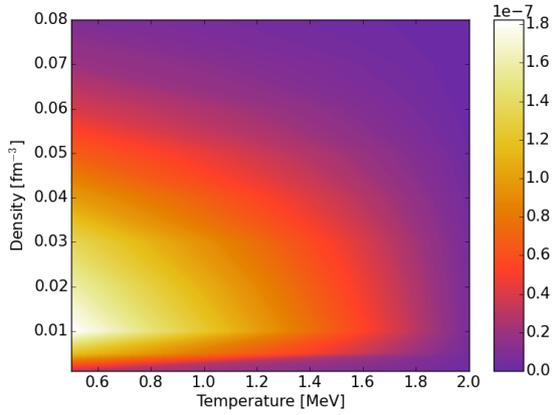
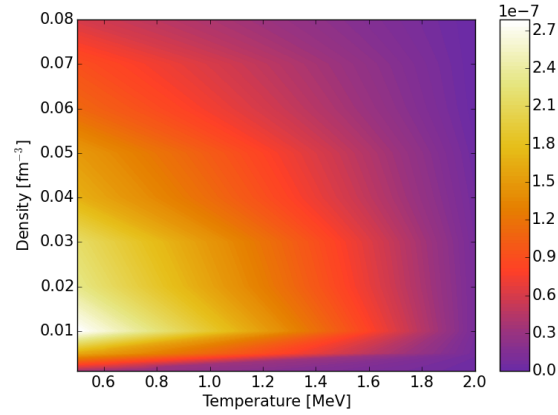
  \centering
  \begin{subfigure}{\columnwidth}
    \includegraphics[width=\columnwidth]{{{s_0.2}}}
    \caption{$x=0.2$}
  \end{subfigure}
  \begin{subfigure}{\columnwidth}
    \includegraphics[width=\columnwidth]{{{s_0.3}}}
    \caption{$x=0.3$}
  \end{subfigure}
  \begin{subfigure}{\columnwidth}
    \includegraphics[width=\columnwidth]{{{s_0.4}}}
    \caption{$x=0.4$}
  \end{subfigure}
  \begin{subfigure}{\columnwidth}
    \includegraphics[width=\columnwidth]{{{s_0.5}}}
    \caption{$x=0.5$}
  \end{subfigure}
  \caption{(Color online) Opacity peak height in the very long
    wavelength for different proton fractions as a function of
    temperature and density. It can be seen that the opacity decreases
    drastically for $T \gtrsim 0.8\,\text{MeV}$. We also show here
    that the opacity is affected by the proton fraction, as it can be
    noted from the scales on the color bar. Also note that in the
    opacity for $x = 0.2$ and $x = 0.3$, the results are governed by
    noise.}
\label{fig:absorption}
\end{figure*}

In figure~\ref{fig:absorption} we show the \emph{opacity peak} for the
different thermodynamic configurations. We can see there that as the
proton fraction decreases, the opacity decreases as well. For every
proton fraction studied, the opacity peak falls rapidly for
temperatures higher than $T=0.8\,\text{MeV}$, and it is about $1/4$ of
the opacity peak at $T=0.5\,\text{MeV}$. The system opacity goes
down as the proton fraction is reduced because the backbone structure
is due to the proton long-range Coulomb interaction. When there is one
neutron for each proton ($x = 0.5$), the neutron structure follows
almost identically that of the proton backbone. However, as the
neutron proportion rises, the neutron structure is smeared out and its
long range correlation begins to vanish. This effect can be seen in
the cluster distribution for $x = 0.2$, where we have many isolated
neutrons, that are the embedding neutron gas. These characteristics
affect the inhomogeneities that appear in $x = 0.5$, suppressing their
long range opacity.

From figure~\ref{fig:large_mass} we can see that even for very high
temperatures ($T = 2.0\,\text{MeV}$) a large cluster appears for every
proton fraction. This large structure is the \emph{Generalized Nuclear
  Pasta}, that is responsible for the long range interaction. The
reason why the opacity gets drastically depressed as the temperature
rises therefore is not because the large cluster disappears, but
because of structural changes.

\section{Discussion and Concluding Remarks}\label{sc:conc}

Neutron rich matter develops non-homogeneous structures (usually
referred to as nuclear pasta) that strongly alter its opacity to
neutrinos. By analyzing the behavior of the neutron-neutron static
structure factor and radial distribution function over a wide range of
densities, temperatures and proton fractions, we are able to calculate
the wavelength at which maximum scattering takes place. We have seen
that at high densities, where very big clusters are expected
(\emph{spaghetti} and \emph{lasagna}), the wavelength stays relatively
constant and the maximum opacity is obtained for rather energetic
neutrinos ($E_\nu \approx 80\,\text{MeV}$, typical of a very early
stage of the evolution of proto-neutron stars). As the density goes
down, we move into the \emph{gnocchi} pasta phase, in which clusters
are of finite size. In this case, the maximum opacity moves to lower
energies. As seen in figure~\ref{fig:absorption} this increase on the
opacity not only takes place when heterogeneities are of the commonly
referred nuclear pasta, but also appears when these structures are
quite deformed (the \emph{generalized nuclear pasta} that we can see
in figure~\ref{fig:morpho}).

We expect these results to be qualitatively correct, but
quantitatively dependent on the model chosen to describe neutron rich
matter. The model we are using in this work has been extensively
studied in collisions and heavy ion physics; that is the reason why we
have chosen it to describe quantitatively neutron rich matter.

Neutron rich matter hydrodynamic models~\cite{ruffert_coalescing_1995,
  mezzacappa_investigation_1998, geppert_temperature_2004,
  woosley_physics_2005, liebendorfer_supernova_2005} can yield proton
fraction, density and temperature for different conditions
(supernovae, proto-neutron stars, neutron stars). From this work, we
are able to find, for this specific model, the neutrino opacity for
different thermodynamic conditions. Therefore, combining these two
results with eventual measurements of the neutrino opacity in
neutron stars, we can check the validity of different nuclear models
and, consequently, move a step forward towards finding the nuclear
equation of state.

\appendix

\section{On the calculation of the structure factor}

The structure factor of a system is defined by the \emph{sample
  scattering amplitude}~\cite{egami_underneath_2003}

\begin{equation}
  \Psi(\mathbf{Q}) = \frac{1}{\langle b\rangle} \sum_i b_i
  \text{e}^{i\mathbf{Q}\cdot\mathbf{R}_i}
  \label{eq:scat_amp}
\end{equation}

with $\mathbf{Q}$ the diffraction vector or momentum
transfer. $\mathbf{R}_i$ is the position of the particle $i$, and
$\langle b\rangle$ is the average of the scattering amplitude of
each particle in the vacuum $b_i$. From this moment on, we will
consider that all of the atoms are of the same species, $b_i = b$.

From $\Psi(\mathbf{Q})$ we define the structure factor
$S(\mathbf{Q})$ as

\begin{equation}
  S(\mathbf{Q}) = \frac{1}{N} |\Psi(\mathbf{Q})|^2
\end{equation}
What follows \emph{immediately} from this expression is that the
structure function must be always positive for every value of
$\mathbf{Q}$. We can expand the scattering amplitude and use $|z| =
z\cdot z^*$ and, if all the atoms are of the same type,

\begin{align}
  S(\mathbf{Q}) &= \frac{1}{N} \left( \sum_i \text{e}^{i\mathbf{Q}\cdot\mathbf{R}_i} \right)
  \left( \sum_j \text{e}^{-i\mathbf{Q}\cdot\mathbf{R}_j} \right)\\
  &= \frac{1}{N} \sum_{i, j} \text{e}^{i\mathbf{Q}\cdot(\mathbf{R}_i-\mathbf{R}_j)}\\
  &= \frac{1}{N} \left[N + \sum_{i < j}
    \left(\text{e}^{i\mathbf{Q}\cdot(\mathbf{R}_i-\mathbf{R}_j)} +
      \text{e}^{i\mathbf{Q}\cdot(\mathbf{R}_i-\mathbf{R}_j)}\right)\right]\\
  &= 1 + \frac{2}{N}\sum_{i < j}\cos{\mathbf{Q}\cdot\mathbf{R}_{ij}}
\end{align}

Usually we are interested in the \emph{powder average} of the structure
factor. This is the structure factor averaged for every possible
orientation of the diffraction vector - because in a powder we have
a lot of structures randomly oriented. We calculate therefore

\begin{equation}
  S(q) = \frac{1}{4\pi}\int\text{d}\phi\text{d}(\cos\theta) S(\mathbf{Q})
\end{equation}

This integral can be performed easily if we put the $z$ axis along
with the direction of $\mathbf{Q}$ and perform the integration
by rotating the distances $\mathbf{R}_{ij}$

\begin{align}
  S(q) &= \frac{1}{4\pi}\int\text{d}\phi\text{d}(\cos\theta)
  \left[1 + 2\sum_{i < j}\cos\left(q\,r_{ij}\,\cos\theta\right)\right]\\
  &= 1 + \frac{1}{2N}\int\text{d}(\cos\theta)
  2\sum_{i < j}\cos\left(q\,r_{ij}\,\cos\theta\right)\\
  &= 1 + \frac{1}{2N} 2 \sum_{i < j} \left.\frac{\sin(q\,r_{ij}u)}{q\,r_{ij}}\right|_{u=-1}^{u=1}\\
  &= 1 + \frac{2}{N} \sum_{i < j}\frac{\sin(q\,r_{ij})}{q\,r_{ij}}
\end{align}

This is the famous Debye formula and, since its the average of an
always positive quantity, it must be always positive.

\begin{figure}  
  \centering
  \includegraphics[width=\columnwidth]{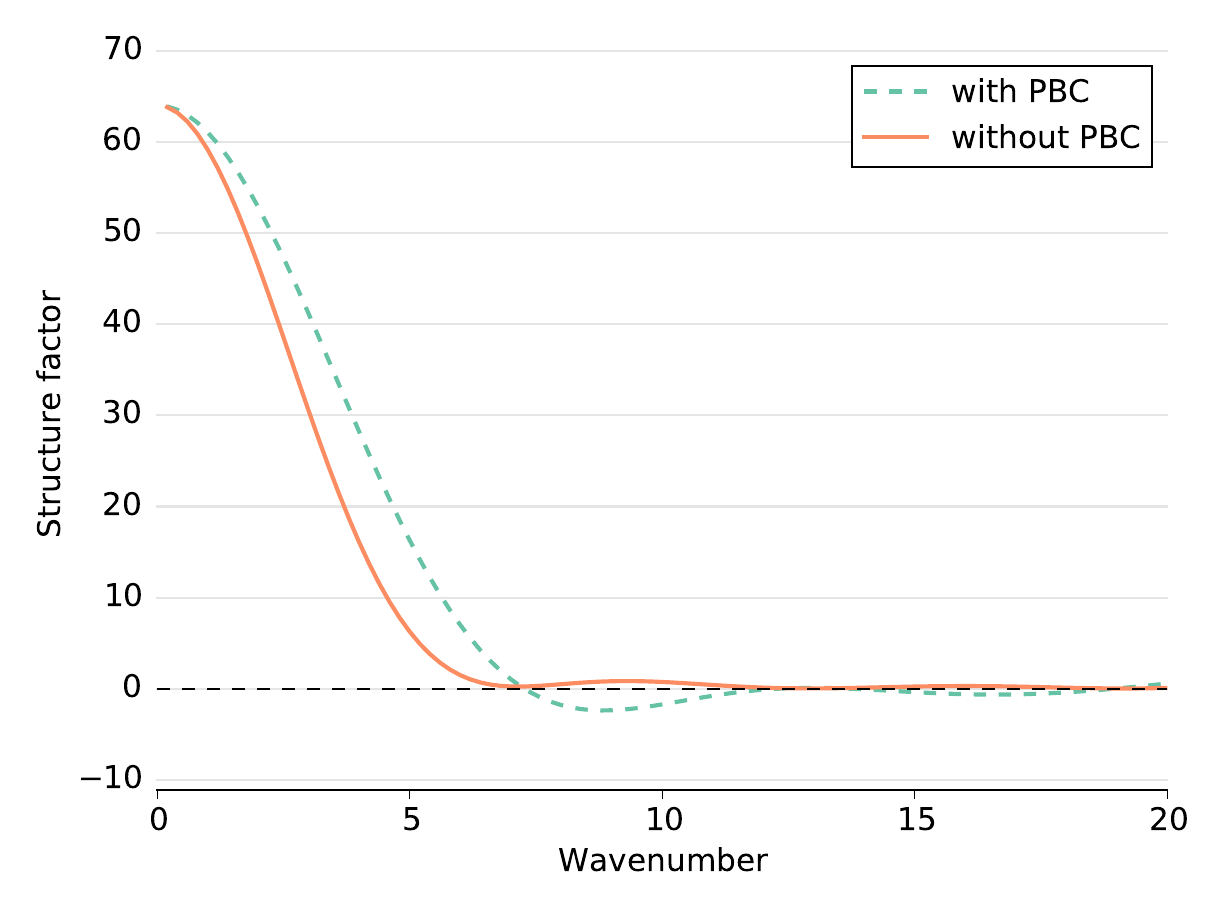}
  \caption{Comparison of structure factor with and without PBC.\@It's
    evident that the structure factor calculated with periodic
    boundary conditions shows negative values, which should not exist
    from the definition of the structure factor.}
\label{fig:ssf_comp}
\end{figure}

One of the most usual problems when we model and study systems in
computer simulations is that we don't have actual \emph{infinite}
systems. We do, however, use the periodic boundary conditions (PBC)
usually to emulate the behavior of infinite systems. With the periodic
boundary conditions we use the minimum image convention: from all the
possible positions through the boundaries for particle $i$ and $j$, we
pick whichever pair is closest. By using the above mentioned method
for a very simple test case (a simple cubic 3D lattice with 4$\times$4$\times$4=64
atoms) we calculated the structure factor that can be seen in
figure~\ref{fig:ssf_comp}.

We can see that the structure factor calculated with PBC attains
negative values, even though those values ought to be forbidden. The
reason for this behavior is the minimum image convention: the pair
distance now isn't always $r_{ij} = r_j - r_i$, but depends on whether
we use the original particles or their images. Therefore, this ``new''
structure factor isn't the product of two complex conjugate 
numbers\footnote{Even further, now the imaginary part of $S(Q)$ is no
  longer zero}. To explore the effect that the minimum image
convention has on the structure factor, we show a comparison of the
structure factor with and without boundary conditions (i.\ e., with the
64 atoms in a void) in figure~\ref{fig:ssf_comp}.

This shows that the structure factor, when we use its definition
\emph{without minimum image convention}, is (as expected) always
positive.

The question then, remains: how can we simulate an infinite medium
when calculating structure factor? The first answer is that it is not
that obvious that we would actually need this \emph{infinite} medium,
since the periodic images of the cell would be aligned in a crystal
that might interfere with the structure within the cell --- the one we
actually do want to study. However, a couple of replicas should be
enough to smear out the finite size effects. One of the possibilities
is to replicate explicitly the box, creating the particles in the
neighboring cells by duplication of the original ones. This, though,
implies a calculation much harder, since the sum is over $N^2$
particles, and replicating only one cell right and left in each
direction would imply a computational time of ${(3^3\cdot N)}^2 \approx
700\cdot N^2$. In general, the complexity $\mathcal{O}(N^2)$ makes
structure factor calculation very expensive for large systems.

There is an alternative to add the boundary conditions. We begin with
the definition of the \emph{sample scattering amplitude} as
in~\ref{eq:scat_amp}, but writing explicitly the periodic boundary
images we want to consider:
\begin{equation}
  \Psi(\mathbf{Q}) = \sum_i \sum_j
  \text{e}^{i\mathbf{Q}\cdot(\mathbf{R}_i+\mathbf{\Delta L}_j)}
\end{equation}
where $\mathbf{\Delta L}_j$ is the distance between a particle and its
$j$-th periodic replica. Since the sums are independent, we can write:
\begin{equation}
  \Psi(\mathbf{Q}) = \left(\sum_i
    \text{e}^{i\mathbf{Q}\cdot\mathbf{R}_i}\right)
  \left(\sum_j\text{e}^{i\mathbf{Q}\cdot\mathbf{\Delta L}_j}\right)
\end{equation}

Multiplying by the conjugate gives us the structure factor
\begin{align}
  S(\mathbf{Q}) &= \left|\sum_i
    \text{e}^{i\mathbf{Q}\cdot\mathbf{R}_i}\right|^2 \left|\sum_j
    \text{e}^{i\mathbf{Q}\cdot\mathbf{\Delta L}_j}\right|^2\\
  &= S_{\text{cell}}(\mathbf{Q})\,S_{\text{PBC}}(\mathbf{Q})
\end{align}

The advantage of this calculation is that it is linear in the sum of
the number of particles $N$ and the number of replicas $M$ consider,
$\mathcal{O}(N+M)$, much lower than the previous
$\mathcal{O}(N^2M^2)$. Consequently, if we want to focus on a region
of $\mathbf{Q}$, this new approach will be useful\footnote{We should
  consider though that in this approach, we will need
  $\mathcal{O}(N+M)$ calculations for each $\mathbf{Q}$, so we can't
  use it to sweep the whole $\mathbf{Q}$ spectrum}. We are left with
only one detail, respecting the \emph{powder average}. It is not
trivial how to calculate this integral, since we need to give proper
weights to each angle. In this work we used the Lebedev
quadrature~\cite{lebedev_values_1975}, although other methods like
Importance Sampling Montecarlo can be useful in this situation.

\section*{Acknowledgments}
This work was partially supported by grants from ANPCyT
(PICT-2013-1692), CONICET and UBACyT. COD acknowledges fruitful
discussions with C Horowitz and F Burgio.

\bibliographystyle{ieeetr}
\bibliography{nuclear}{}
\end{document}